\documentclass[12pt]{article}

\catcode`\@=11
\@addtoreset{equation}{section}

\global\arraycolsep=2pt
\usepackage{mathrsfs,amsbsy,amssymb,latexsym,amsfonts,amsmath,cite}
\usepackage{graphicx,color}

\allowdisplaybreaks

\newcommand{\bea}{\begin{eqnarray}}
\newcommand{\eea}{\end{eqnarray}}
\newcommand{\beq}{\begin{equation}}
\newcommand{\eeq}{\end{equation}}

\def\eqref#1{Eq.~(\ref{#1})}

\newcommand{\pd}{\partial}

\def\X5sp{{\rm X}_5}
\def\Y3sp{{\rm Y}_3}
\def\Z3sp{{\rm Z}_3}

\def\e{{\rm e}}

\begin{document}

\begin{flushright}
\parbox{4cm}
{KUNS-2436}
\end{flushright}

\vspace*{0.5cm}

\begin{center}
{\Large \bf  Dynamical Lifshitz-type solutions \\ and aging phenomena
}
\vspace*{1.5cm}\\
{\large Kunihito Uzawa$^{\ast}$\footnote{E-mail:~uzawa@yukawa.kyoto-u.ac.jp}  
and Kentaroh Yoshida$^{\dagger}$\footnote{E-mail:~kyoshida@gauge.scphys.kyoto-u.ac.jp}} 
\end{center}
\vspace*{0.25cm}
\begin{center}
$^{\ast}${\it Department of Physics,
School of Science and Technology, \\ 
Kwansei Gakuin University, Sanda, Hyogo 669-1337, Japan
}%
\vspace*{0.5cm} \\ 
$^{\dagger}${\it Department of Physics, Kyoto University \\ 
Kyoto 606-8502, Japan 
} 
\end{center}
\vspace{1cm}
\begin{abstract}
We consider time-dependent Lifshitz-type solutions in type IIB supergravity. 
The solutions describe a time evolution from Lifshitz spacetimes to AdS 
spaces. We argue the holographic relation of them to aging phenomena in 
condensed matter physics. The solutions have no time-translation invariance 
and possess the dynamical scaling symmetry with $z_c=2$\,. 
In addition, the time evolution corresponds to slow (nonexponential) 
dynamics from nonequilibrium states to the equilibrium. We also discuss 
a mechanism of quantum quench to generate nonequilibrium states.   
\end{abstract}

\setcounter{footnote}{0}
\setcounter{page}{0}
\thispagestyle{empty}

\newpage

\section{Introduction}

The AdS/CFT correspondence \cite{M,GKP,W} is 
supported by an enormous amount of evidence to date, and 
there is no doubt about its validity, at least in the planar limit.  
An attractive direction in the study of AdS/CFT is to consider 
its applications to realistic systems such as QCD \cite{PSS,KSS,SS} 
and condensed matter systems \cite{RT,HHH,Son,BM,Kachru}. 
(For reviews, for example, see \cite{Lecture1,Lecture2,Lecture3}). 

Our purpose here is to consider holographic descriptions of nonequilibrium 
states dynamically relaxing to the equilibrium state. Since AdS spaces 
correspond to equilibrium states, the gravitational solutions should 
be time dependent and describe a time evolution to AdS spaces.  
However, it is generally difficult to consider time-dependent AdS/CFT.  
It is of great importance to understand how to describe nonequilibrium 
systems holographically in order to obtain the deeper understanding of 
gravity and AdS/CFT itself. 

In this paper we construct time-dependent Lifshitz-type exact solutions 
in type IIB supergravity and argue their holographic descriptions. 
We show some evidence to support that the solutions describe 
a class of nonequilibrium phenomena called the aging 
\cite{Struik,glassy,HP}, which are observed, for example, 
in glassy materials.     
The phenomena are characterized by the following three features: 
(1) no time-translation invariance, (2) the dynamical scaling with 
the dynamical critical exponent $z_c$ and (3) slow (nonexponential) 
relaxation. The solutions we construct here satisfy all of them. 
For earlier holographic approaches to the aging, see \cite{Minic,Leigh}.   

One of the techniques to prepare nonequilibrium states is quenching 
the system. There are various ways of quenching. Since we consider 
the zero-temperature solutions, a quantum quench, which suddenly changes 
a coupling constant, plays a central role our analysis. 
We also explain a mechanism 
of the quantum quench encoded into the gravitational solutions. 

Another aspect of the subject is its potential application to realistic 
cosmological models. The construction of the time-dependent Lifshitz-type 
solutions is similar to that of dynamical brane solutions 
\cite{GLP, CCGLP, KU1}, which provide the Friedmann-Robertson-Walker 
universe. Indeed, a dynamical universe is also obtained from the 
Lifshitz-type solutions, though there are some undesirable aspects.

\smallskip 

\section{Dynamical Lifshitz-type solutions}

Let us consider D3-brane solutions with waves in type IIB supergravity. 
We take account of the metric $g_{MN}$, dilaton $\phi$, axion $\chi$ 
and the self-dual five-form field strength $F_{(5)}$.  
Then the field equations are given by
\begin{eqnarray}
&& R_{MN} = \frac{1}{2}\pd_M\phi\pd_N\phi
+\frac{1}{2}\e^{2\phi}F_{M} {F_N} 
+\frac{1}{4\cdot 4!}F_{MA_2\cdots A_5} 
{F_N}^{A_2\cdots A_5}\,,
\nonumber \\ 
&& F_{(1)} \equiv d\chi\,, \quad 
d\left[\e^{2\phi}\ast F_{(1)}\right]=0\,, \quad 
F_{(5)}=\ast F_{(5)}\,, \nonumber  
\end{eqnarray}
where $\ast$ is the Hodge operator in ten dimensions. 

We are concerned with a time-dependent generalization of  $z_c=0$ 
Schr\"odinger spacetimes 
in type IIB supergravity. Hence the following ansatz is supposed,   
\begin{equation}
ds^2 = h^{-1/2}(r) \Bigl[ 2dudv + G(u ,v, r)du^2 
+ \sum_{i=1}^2(dy^i)^2 \Bigr] 
 + h^{1/2}(r) \Bigl[dr^2+r^2d\Omega_{(5)}^2\Bigr]\,, \label{general-sol}
\end{equation}
which is equipped with 
\begin{equation}
F_{\left(5\right)}=(1\pm\ast) d\left[h^{-1}(r)\wedge du
\wedge dv\wedge dy^1\wedge dy^2\right]\,, \quad 
\phi=\phi_0\,, \quad F_{\left(1\right)}=k\,du\,, \nonumber 
\end{equation}
where $\phi_0$, $k$ are constants and 
$d\Omega_{(5)}^2$ describes the five-dimensional sphere with unit radius. 
The scalar functions $G(u, v, r)$ and $h(r)$ are not determined yet.  
The general forms of them are given by, respectively,  
\begin{equation}
G(u, v, r) = c_0 + c_1\, u + c_2\, v + \frac{c_3}{r^4} 
+\frac{k^2}{4}\e^{2\phi_0}\left(\frac{L^4}{r^2}-\frac{r^2}{3}\right), 
\quad 
h(r) = 1 + \frac{L^4}{r^4}\,.  \nonumber 
\end{equation}
Here $c_a~(a=0,\ldots,3)$ and $L$ are constant parameters. 
When $c_1=c_2=0$\,, (\ref{general-sol}) is reduced to the general 
solutions preserving the $z_c=0$ Schr\"odinger symmetry 
\cite{CH}. Nonvanishing $c_1$ and $c_2$ break time-translation 
invariance and phase-rotation symmetry, respectively. 
Note that $c_0$ can always be removed by shifting $v$ as $v \to v - 
\frac{1}{2}c_0\,u$\,, up to the redefinition of $c_1$\,. 
Henceforth, we will set $c_0 = 0$. 

Let us consider the solution (\ref{general-sol}) with $c_1=c_3=0$\,, 
for simplicity. Then the near-horizon limit leads to the following metric,  
\begin{equation}
ds^2 = \frac{r^2}{L^2} \Bigl[ 2dudv + F(r,v)
\frac{du^2}{r^2} + (dy^1)^2 +(dy^2)^2  \Bigr] 
+ \frac{L^2}{r^2}dr^2 + L^2d\Omega_{(5)}^2\,, 
\label{Schrodinger} 
\end{equation}
where we have introduced the following quantities, 
\begin{equation}
F(r,v) \equiv c_2\, r^2\, v + \frac{k^2g_s^2L^4}{4}\,, \qquad g_s \equiv 
{\rm e}^{\phi_0}\,. \nonumber 
\end{equation}
Here $g_s$ is the string coupling constant. It is necessary to take the 
string coupling $g_s \ll 1$ so as to ignore graviton loop effects. 
This condition is ensured by taking the scaling limit, 
\[
k \to \infty\,, \quad g_s \to 0\,, \qquad kg_s~:\mbox{fixed}\,,
\]
as well as the usual scaling limit, 
\[
N \to \infty\,, \quad g_s \to 0\,, \qquad Ng_s~:\mbox{fixed}\,,
\]
where $N$ is the number of D3-branes. 
Note that $Ng_s$ and $kg_s$ are fixed to be two independent constants. 

When $c_2=0$\,, the metric (\ref{Schrodinger}) is the direct product of 
the $z_c=0$ Schr\"odinger spacetime  
and the five-dimensional sphere S$^5$ with the unit radius. 
However, the constant shift symmetry for $v$ is broken when $c_2 \neq 0$\,. 
This means that the nonvanishing $c_2$ breaks the phase rotation 
in the Schr\"odinger algebra with $z_c=0$\,. 

In summary, the metric (\ref{Schrodinger}) is invariant under (i) a 
constant shift of $u$\,, (ii) spatial translations of $y^i$~(i=1, 2), 
(iii) a rotation on the $y^1$-$y^2$ plane, (iv) Galilean symmetries 
and (v) the scale transformation with $z_c=0$\,, 
\begin{eqnarray}
 u \to u\,, \quad v \to \lambda^2\,v\,, \quad y^i \to \lambda\,y^i\,, \quad 
 r \to \frac{1}{\lambda}\,r\,, \nonumber 
\end{eqnarray} 
where $\lambda$ is a constant parameter. 

The next step is to derive time-dependent Lifshitz-type solutions 
by applying the trick in \cite{DG} to the solutions (\ref{Schrodinger})\,. 

The metric (\ref{Schrodinger}) is first rewritten as 
\begin{equation}
ds^2 = - \frac{r^4}{L^2\,F(r,v)} dv^2 
+ \frac{r^2}{L^2} (dy^i)^2 + L^2 \frac{dr^2}{r^2} 
+ \frac{F(r,v)}{L^2} \left(du + \frac{r^2}{F(r,v)} dv\right)^2 + 
L^2 d\Omega_{(5)}^2\,. \label{Lifshitz}
\end{equation}
Since the $u$ direction is still invariant under a constant shift, 
one may impose a periodic boundary condition for this direction.

The boundary condition breaks the Galilean symmetries as shown in 
\cite{BN,DG}. Now that the coordinate $v$ can be regarded as the time 
direction rather than $u$, 
the geometry should be understood as a time-dependent Lifshitz spacetime. 
The remaining symmetry is close to the aging algebra \cite{aging-symm}, 
but special conformal and Galilean symmetries are not contained. 

Note that a problem happens in the Kaluza-Klein (KK) reduction. 
This is an intrinsic point to the dynamical case.  
The radius of compactification is not small everywhere in comparison to 
the characteristic length-scale $L$ of the solutions. 
To make matters worse, it even shrinks to zero when $F(r,v)=0$\,. 
Hence, the compactification produces another curvature singularity, 
apart from divergent tidal forces intrinsic to Lifshitz and Schr\"odinger 
spacetime \cite{Lecture1,Kachru,pathology1,pathology2}. 
Indeed, when $F(r,v)=0$\,, the KK-gauge field also diverges.

\section{The behavior of the solutions} 
To see the time evolution of the solutions (\ref{Lifshitz}), 
we use the following coordinates: 
\begin{eqnarray}
r = \frac{L^2}{z}\,, \qquad v = \frac{kg_s}{2} \tau\,. \nonumber 
\end{eqnarray}  
Then the metric (\ref{Lifshitz}) is rewritten as 
\begin{eqnarray}
ds^2 &=& \frac{L^2}{z^2}\left[ -\frac{d\tau^2}{z^2 f(z,\tau)} 
+ (dy^i)^2 + dz^2 \right] 
+ L^2 d\Omega_{(5)}^2 \nonumber \\ 
&&  
+ \frac{k^2 g_s^2}{4}f(z,\tau)L^2\left( du 
+ \frac{2}{kg_s z^2 f(z,\tau)}d\tau\right)^2\,, 
\label{relaxation}
\end{eqnarray}
where we have defined the following quantities:
\[
f(z,\tau)  \equiv 1 + \frac{2\tau}{k^2g_s^2z^2 \tau_0}\,, 
\qquad c_2 \equiv \frac{1}{kg_s\tau_0}\,. 
\]
We assume that $c_2>0$ hereafter and $\tau_0$ is later 
regarded as a microscopic time scale.  

The new metric (\ref{relaxation}) describes the Lifshitz spacetime when 
$f(z,\tau) \simeq 1$. Hence, the Lifshitz spacetime dominates for all 
of the values of $z>0$, when $\tau \simeq 0$. 

\begin{figure}[htbp]
\begin{center}
\includegraphics[scale=.34]{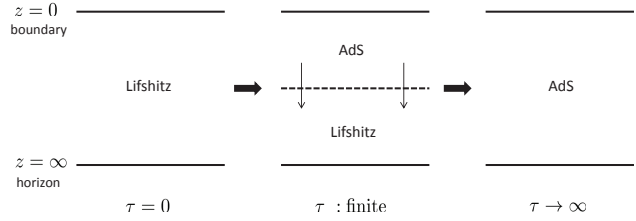}
\end{center}
\vspace*{-0.5cm}
\caption{Time evolution of the dynamical Lifshitz solutions.}
\label{evolution:fig}
\end{figure}

We first consider the time evolution for $\tau >0$. As time progresses, 
the metric (\ref{relaxation}) tends to deviate from the Lifshitz at 
$\tau =0$. It is convenient to divide the spacetime into the following 
two regions at a fixed time $\tau$: 
\begin{eqnarray}
&& \mbox{(i)} \quad \mbox{the near-boundary region} \quad z^2 \ll 
\frac{1}{k^2g_s^2}\frac{\tau}{\tau_0}\,, \nonumber \\ 
&& \mbox{(ii)} \quad \mbox{the near-horizon region} \quad z^2 \gg 
\frac{1}{k^2g_s^2}\frac{\tau}{\tau_0}\,. \nonumber 
\end{eqnarray} 
In the near-horizon region, it still remains the Lifshitz spacetime. 
To evaluate the behavior of the metric in the near-boundary region, 
we take the $\tau \to \infty$ limit. Then 1 in $f(z,\tau)$ is ignored. 
By performing the coordinate transformation, 
\begin{equation}
\tau = \frac{\tau'{}^2}{2k^2g_s^2\tau_0}\,, \label{cosmic}
\end{equation}
the metric  (\ref{relaxation}) is rewritten as  
\begin{equation}
ds^2 = \frac{L^2}{z^2}\left[ -d\tau'{}^2 + (dy^i)^2 + dz^2 \right] 
+ L^2 d\Omega_{(5)}^2 
+ \left(\frac{L\tau'}{2kg_s z \tau_0}\right)^2\left( du 
+ \frac{2kg_s \tau_0}{\tau'}d\tau'\right)^2\,. \nonumber 
\end{equation}
Thus the usual AdS metric has been reproduced.   
This implies that the Lifshitz spacetime at $\tau=0$ tends to decay to the 
AdS space from the boundary as $\tau \to \infty$\,. The time evolution is 
shown in Fig.~\ref{evolution:fig}. 
Note that the compactification radius is large enough in the near-boundary 
region or late time and, hence, the KK direction is effectively 
decompactified in the AdS region. 

The next issue is to study the conformal boundary of the solutions 
(\ref{relaxation})\,. It is helpful to go back to the coordinate $r$. 
Then the metric is expressed as 
\begin{eqnarray}
&&ds^2 = \frac{r^2}{L^2}\left[- 
\frac{k^2g_s^2}{2g(r,\tau)}\left(\frac{\tau_0}{\tau}\right)d\tau^2 
+(dy^i)^2\right] + \frac{L^2}{r^2} dr^2  \nonumber \\ 
\nonumber \\ 
&&~~~~~~~~+ L^2 d\Omega_{(5)}^2 + \frac{r^2}{2L^2}
\left(\frac{\tau}{\tau_0}\right)g(r, \tau)
\left(du + \frac{kg_s}{g(r, \tau)}\left(\frac{\tau_0}
{\tau}\right)d\tau\right)^2, \nonumber 
\end{eqnarray}
where we have defined a new function, 
\[
g(r,\tau)  \equiv 1 + \frac{k^2g_s^2 L^4\tau_0}{2r^2 \tau}\,.
\]
Then by performing a conformal transformation for the metric as 
$ds^2 \to \Omega^2(r,\tau) ds^2$ with the conformal factor, 
\begin{eqnarray}
\Omega^2(r,\tau) \equiv \frac{L^2}{r^2}\frac{2g(r,\tau)}{k^2g_s^2}
\frac{\tau}{\tau_0} \qquad (\tau >0)\,, \label{factor}
\end{eqnarray}
and by taking the limit $r\to \infty$, 
the metric becomes 
\begin{equation}
ds^2 = -d\tau^2 + 
\frac{2}{k^2g_s^2}\,\left(\frac{\tau}{\tau_0}\right)\,(dy^i)^2 
+ \frac{1}{k^2g_s^2}
\left(\frac{\tau}{\tau_0}\right)^2\left(du + 
kg_s\left(\frac{\tau_0}{\tau}\right)d\tau \right)^2 
\nonumber 
\end{equation}
and describes a dynamical universe with a power-law expansion. 
It is notable that the conformal boundary is well defined, differently 
from the Lifshitz spacetime. 

One may think that there is an external force to drive the 
expansion of the universe. However, the time dependence in $y^i$ 
directions comes from the time-dependent conformal factor (\ref{factor}). 
It makes the length scale on the boundary time dependent. This situation 
is described as the dynamical universe. Thus, the expansion is superficial. 
Indeed, the time dependence disappears by taking $\tau'$ introduced 
in (\ref{cosmic}). Then the time-independent conformal factor leads to 
Minkowski spacetime, up to the dynamical S$^1$ part. 

\begin{figure}[htbp]
\vspace*{-0.35cm}
\begin{center}
\includegraphics[scale=.23]{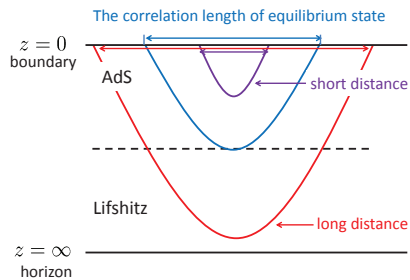}
\end{center}
\vspace*{-0.85cm}
\caption{The short-distance physics is dominated by the AdS space, 
while the long-distance physics 
is influenced by the Lifshitz spacetime. The correlation length of 
equilibrium state is estimated 
with the middle point between the two regions. 
}
\label{length:fig}
\end{figure}

Let us argue what happens in the boundary theory from the bulk structure. 
Now the AdS/CFT dictionary tells us that the short-distance physics is 
dominated by the bulk AdS space corresponding to equilibrium states. 
On the other hand, the Lifshitz spacetime influences the long-distance 
physics, and it may be regarded as nonequilibrium states 
(See Fig.~\ref{length:fig}). 
This means that there exists no long-range order, only ordered structure 
like the crystalline lattice at short distances, such as that found in 
the materials that exhibit the aging phenomena. 

The correlation length of equilibrium state depends on time because of 
the expansion of the universe. The time dependence is evaluated as
\begin{eqnarray}
L(\tau) = \left(\frac{\tau}{\tau_0}\right)^{1/2}L(\tau_0)\,, \label{length}
\end{eqnarray}
where $L(\tau_0)$ can be evaluated holographically by using the middle 
point value of the $z$ direction, $z_{\rm mid} = 1/kg_s$, 
as in Fig.~\ref{length:fig}. This time dependence agrees with $L(\tau) 
\sim \tau^{1/z_c}$ in the aging phenomena with $z_c=2$\,. 

Now the physical meaning of $\tau_0$ is also obvious. When 
$\tau_0 = \infty$, $L(\tau)$ vanishes and, hence, no relaxation occurs. 
The whole geometry of the bulk is the Lifshitz spacetime. 
When $\tau_0=0$,  $L(\tau) =\infty$ becomes infinite, 
and the system is completely relaxed. Then the whole bulk is the AdS space. 
Thus, $\tau_0$ should be identified with a typical relaxation time at 
microscopic scale.   

\vspace*{-0.2cm}

\begin{figure}[htbp]
\vspace*{-0.4cm}
\begin{center}
\includegraphics[scale=.34]{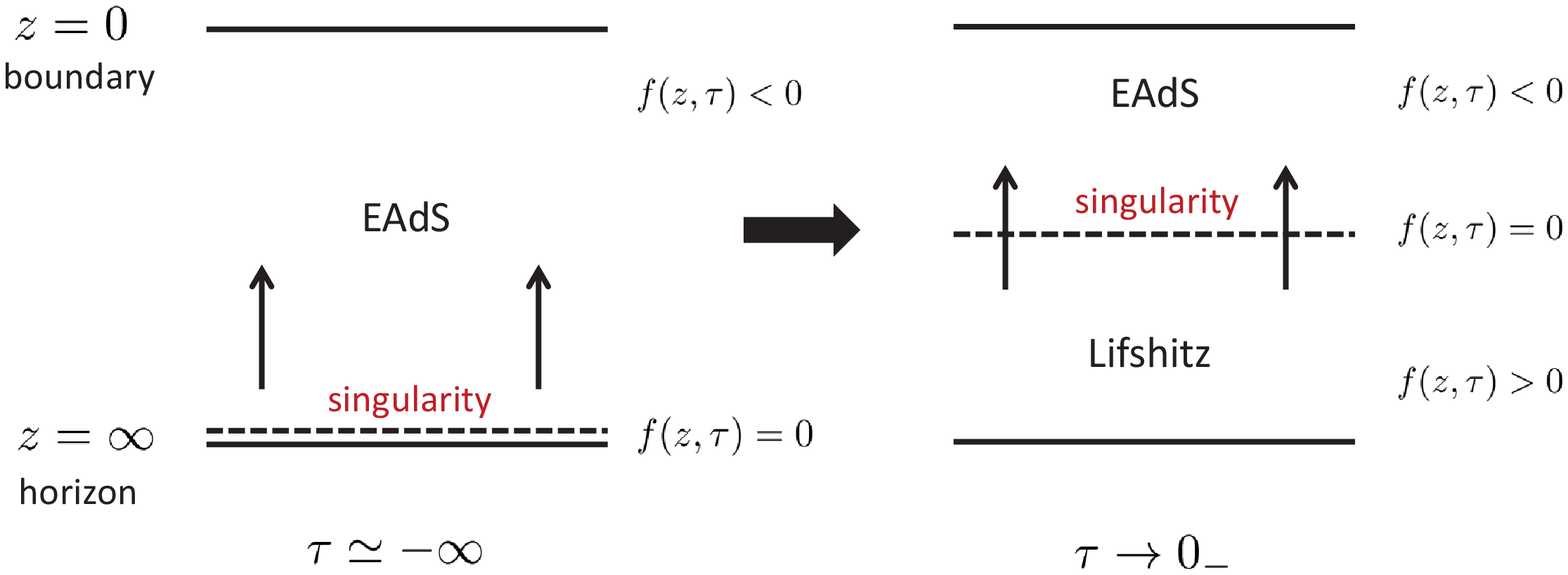}
\end{center}
\vspace*{-0.78cm}
\caption{A moving curvature singularity when $\tau < 0$. }
\label{bump:fig}
\vspace*{-0.25cm}
\end{figure}

Finally let us discuss the behavior of the solutions (\ref{relaxation}) 
when $\tau <0$\,. There is a curvature singularity when $f(z,\tau)=0$. 
The location of the singularity is  
\begin{eqnarray}
z_s(\tau) \equiv \frac{\sqrt{2}}{kg_s}\sqrt{-\frac{\tau}{\tau_0}}    
\end{eqnarray} 
and depends on time $\tau$\,. It appears from the horizon at 
$\tau = -\infty$, and it goes up to the boundary 
and finally runs into the boundary at $\tau\to 0$\,. 
The time evolution for $\tau <0$ is depicted in Fig.~\ref{bump:fig}. 
In the region where $f(z,\tau) <0$\,, the $\tau$ coordinate cannot be 
regarded as the time direction any more. The metric in this region 
behaves as a Euclidean AdS (EAdS) space and the $u$ direction becomes 
timelike. Now that the $u$ direction is compactified, a closed timelike 
curve exists in this region. 

The bump of the singularity may be interpreted as a quantum quench 
to produce nonequilibrium states. The effective time-dependent coupling 
$g_{\rm eff}(\tau)$, defined as 
\begin{eqnarray}
kg_{\rm eff}(\tau) \equiv kg_s \sqrt{\frac{\tau_0}{|\tau|}}\,, 
\end{eqnarray} 
diverges rapidly as $\tau \to 0$, and this may be understood as a 
holographic description of quantum quench \cite{quench}. 


In conclusion, we have shown time-dependent Lifshitz-type solutions in 
type IIB supergravity and discussed the time evolution in relation to 
the aging phenomena. It is nice to look for more general solutions 
corresponding to more complicated processes.  
A finite-temperature generalization is especially interesting. 

As another aspect, our solutions give rise to a dynamical universe with a 
power-law expansion. This universe is not isotropic because of the KK circle 
expanding with time, and hence it would be difficult to construct realistic 
models in the present setup. It is interesting to look for other setups 
applicable to building the realistic model. 

We hope that our results develop a new frontier for studying the 
time-dependent AdS/CFT correspondence and find many applications to 
nonequilibrium phenomena in condensed matter physics. 

We would like to thank S.~Nakamura for useful discussions. 
The work of K. Y. was supported by the scientific grants from the Ministry 
of Education, Culture, Sports, Science 
and Technology (MEXT) of Japan (No.\,22740160). 
This work was also supported in part by the Grant-in-Aid 
for the Global COE Program, ``The Next Generation of Physics, Spun 
from Universality and Emergence'', from MEXT, Japan.

\end{document}